\documentclass[journal]{IEEEtran} 

\ifCLASSINFOpdf

\else

\fi

\usepackage[cmex10]{amsmath}
\usepackage[noadjust]{cite}

\usepackage{gensymb}
\usepackage[top=1in, bottom=1in, left=1in, right=1in]{geometry}
\usepackage{graphicx}
\usepackage[colorinlistoftodos]{todonotes}
\usepackage{hyperref}
\usepackage{marginnote}

\usepackage{graphicx}
\usepackage{dcolumn}
\usepackage{bm}
\usepackage{gensymb}
\usepackage{soul}

\newcommand{\RN}[1]{%
  \textup{\uppercase\expandafter{\romannumeral#1}}%
}

\usepackage{soul}
\usepackage[english]{babel}
\usepackage[utf8]{inputenc}
\begin{document}

%
%
\title{Flow sensor based on the snap-through detection of a curved micromechanical beam}

\author{Yoav Kessler,  B. Robert Ilic, Slava Krylov,~\IEEEmembership{Member,~IEEE}, Alex Liberzon}

\maketitle

\begin{abstract}
We report on a flow velocity measurement technique based on snap-through detection of an electrostatically actuated, bistable micromechanical beam. We show that induced elecro-thermal Joule heating and the convective air cooling change the beam curvature and consequently the critical snap-through voltage ($V_{ST}$). Using single crystal silicon beams, we demonstrate the snap-through voltage to flow velocity sensitivity of $dV_{\text{ST}}/du \approx0.13$~V\,s\,m$^{-1}$ with a power consumption of $\approx360\; \mu$W. Our experimental results were in accord with the reduced order, coupled, thermo-electro-mechanical model prediction. We anticipate that electrostatically induced snap-through in curved, micromechanical beams will open new directions for the design and implementation of downscaled flow sensors for autonomous applications and environmental sensors.
\end{abstract}

\begin{IEEEkeywords}
MEMS, snap-through, curved microbeam, flow sensor, electrothermal actuator
\end{IEEEkeywords}

\IEEEpeerreviewmaketitle
\section{Introduction}

\IEEEPARstart{F}{low} sensors based on microelectromechanical systems (MEMS) are attractive due to their small size, low power consumption, high sensitivity, and compatibility with electronic device integration~\cite{ho1998micro,lofdahl1999mems, wang2009mems}. The MEMS-based flow sensors that have been developed in past years~\cite{wang2009mems,gad1989advances} operate in either thermal or non-thermal mode. Thermal flow detectors are based on calorimetry or hot-wire~\cite{borisenkov2015} sensing whereas non-thermal devices are based on force sensing~\cite{svedin2003new}.

Recently, we demonstrated a flow sensor based on a straight, double-clamped, micromechanical beam that buckles under an electro-thermally induced, compressive axial force and convective air flow cooling ~\cite{kessler2016flow}. Flow velocity was  obtained by measuring either the Joule heating current through the beam at the critical buckling or the post-buckling deflection of the beam. Here we present a {gas} flow sensor based on a double-clamped, single crystal silicon beam with lithographically defined in-plane curvature. In this scenario, the snap-through (ST) instability is induced by electrostatic forces, {while Joule heating is used for fine tuning of the beam's curvature near the ST point}.

\begin{figure}[t]
\centering
\includegraphics[width=0.8\columnwidth]{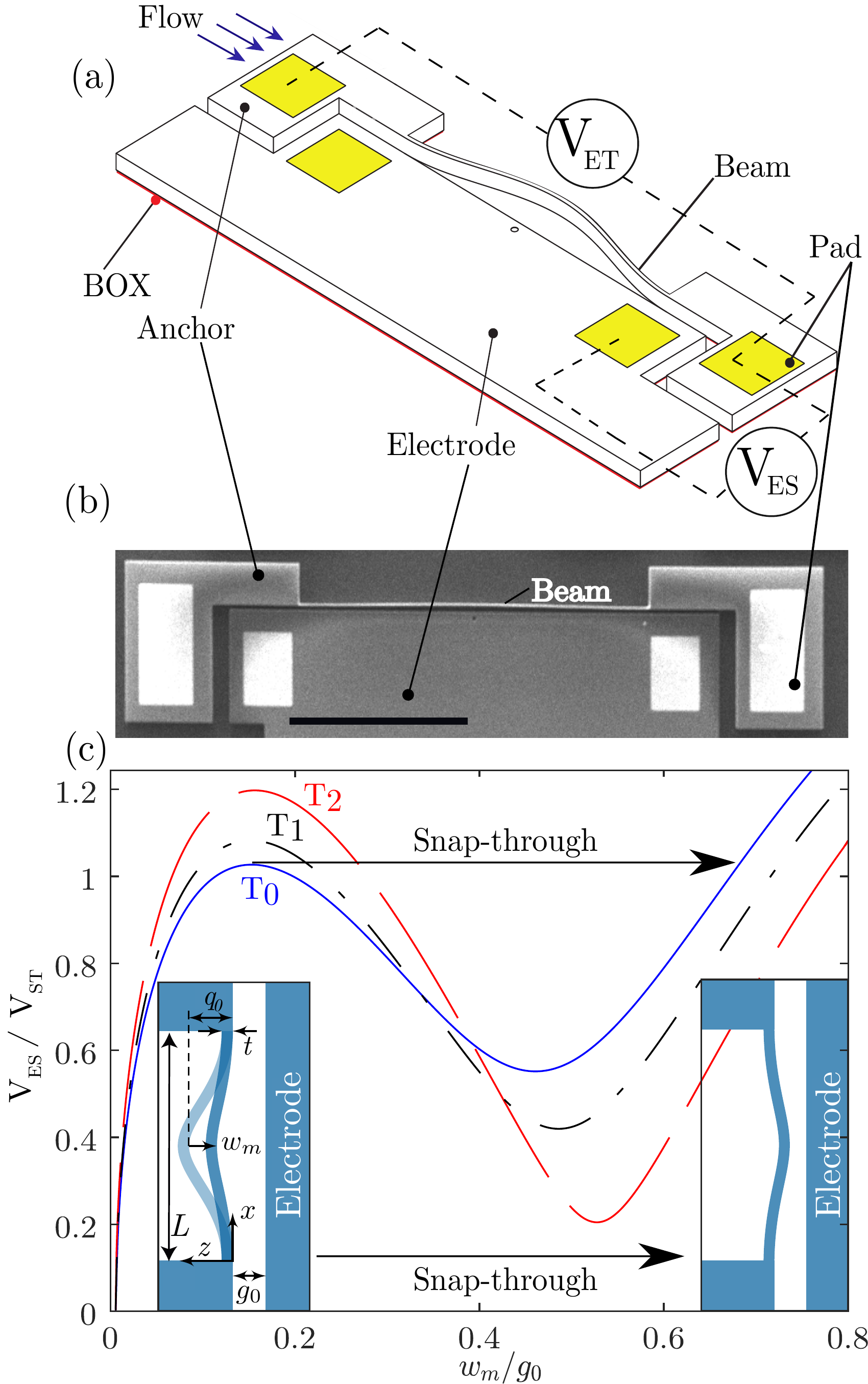}
\caption{(a)~Schematic illustration of the micromechanical flow sensor. The double-clamped, curved beam is electrostatically actuated using a parallel-plate electrode configuration. Heating and cooling of the microstructure is accomplished by the induced current through the beam and the air flow across the structure, respectively. (b)~A top-down scanning electron micrograph of the fabricated device. The scale bar corresponds to 0.5 mm. (c)~Modelling results showing limit point buckling curves of the beam at different temperatures ($T_2> T_1> T_0$). The midpoint deflection {$w_m=q_0-q$ (where $q_0$ and $q$ are the midpoint elevations in the initial and deformed states, respectively)} is normalized by the distance $g_0$ between the electrode and the beam's ends. The actuating voltage $V_{\text{ES}}$, normalized by the snap-through voltage $V_{\text{ST}}$, corresponds to the reference ambient temperature $T_0$. The arrow represents the snap-through collapse. The insets depict the geometry of the beam in its initial, as fabricated, state, in a configuration prior to the ST buckling (left) and post-buckled configuration (right).\label{fig:scheme}}
\end{figure}

Figure~\ref{fig:scheme} shows the micromechanical flow sensor comprising of a curved beam with length $L\approx1000\,\mu$m, width $b\approx20\,\mu$m, thickness $t\approx 2.6\,\mu$m and initial midpoint elevation $q_0\approx1.5\,\mu$m. The device is electrostatically actuated by applying a voltage $V_\text{ES}$ between the beam and a stationary electrode, located at a distance $g_0\approx10.3\,\mu$m from the beam's ends. The voltage difference $V_\text{ET}$ applied between the two anchors induces an electric current through the beam that consequently results in the electro-thermal, Joule heating of the device. {The air flow across the device cools the beam.}

Figure~\ref{fig:scheme}(c), illustrating the device operation, shows typical voltage-deflection characteristics of the beam held at three different temperatures $T_0<T_1<T_2$. When $V_\text{ES}$ exceeds the critical snap-through value $V_\text{ST}$ the curved, bistable beam jumps towards the second stable state. The sensitivity of $V_\text{ST}$ to temperature governs the device functionality. Specifically, at a higher temperature ($T_2$ in Fig.~\ref{fig:scheme}c), due to the compressive axial thermal stress, both the beam curvature and consequently the distance to the electrode increase. This change results in a higher snap-through voltage than the baseline, ambient temperature, value. On the other hand, when the air {flows, the beam is cooled from} $T_2$ to $T_1<T_2$ reduces the the axial stresses within the beam, which results in a lower midpoint elevation and a lower snap-through voltage. Therefore, a measurement of $V_\text{ST}$ provides a direct insight into the air-flow velocity.

\section{Model and methods}
{In this work, we consider only the static response of the device.} The equilibrium of the beam is governed by the equation~\cite{Medina2014323}
\begin{multline}\label{EQ:beamDef}
{E} {I}_{{y}{y}} \left( {z}''''-{z}_0'''' \right) - \left[ {N} + \frac{{E} {A}}{2\,{L}} \int_0^L \left( {z}'^2-{z}_0'^2 \right)d{x}  \right] {z}'' \\ = -\frac{\varepsilon_0 b\, {V^2_\text{ES}}}{2({g}_0+{z}(x))^2}
\end{multline}
\noindent where $z(x)$ and $z_0(x)$ are, respectively, the deformed and nominal, as-designed, elevations of the beam above its anchored ends, $E=169$~GPa is the Young's modulus of Si in the $\langle 110 \rangle$ direction
~\cite{glassbrenner1964thermal, okada1984precise}.
${A}= b t $ and ${I}_{yy}= b t^3/12$ are the area, and the second moment of area of the beam cross section, respectively, and  $(\;)'\equiv d/d{x}$ denotes the derivative with respect to the coordinate  $x$ along the beam. The right hand side of Eq.~\eqref{EQ:beamDef} (where $\varepsilon_0~=~8.85 \times 10^{-12}$~F/m is the permittivity of vacuum)  represents the electrostatic force acting on parallel capacitor plates.

The axial force
\begin{equation}
{N}={\sigma}_{r}{A}-{\alpha}\,\it \overline{\theta}{E}{A}
\label{EQ:N0}
\end{equation}
\noindent (positive when tensile) is engendered by the residual and thermal stresses. In. Eq.~(\ref{EQ:N0}), $\sigma_{r}$ is the residual  stress, {$\alpha~=~3.28\times 10^{-6}~^{\circ}$C$^{-1}$} is the coefficient of thermal expansion of Si and $\overline{\theta} = \frac{1}{L} \int_0^L  \left(T_b(x) - T_\infty \right) dx$ is the mean temperature difference between the beam temperature, $T_b$, {calculated using the one-dimensional heat transfer equation}~\cite{Holman:1986,kessler2016flow}, and the {ambient} temperature of the flow, $T_\infty$.

Using Galerkin decomposition, we set $z(x)=q \, \phi(x)$, $z_0(x)=q_0 \, \phi(x)$, where $\phi(x)$ is the first buckling mode of a straight beam~\cite{timoshenko2009theory} and $q$ is the midpoint  elevation of the deformed beam above its ends, and obtain
\begin{multline}
w_{{m}} \left( 1 +\frac{N}{N_E}+\frac{{q_{0}}^{2} }{8r^2}\right)-  3\frac{{w_m}^{2}q_{0} }{16r^2}   + \frac{{w_m}^{3}}{16r^2} \\
 =\frac{N}{N_E}q_{{0}}+{\frac {\beta}{   \left(
 g_0+q_{{0}}-w_{{m}}  \right) ^{3/2}}}\label{EQ:beamRO}
\end{multline}
\noindent Here ${\beta}~=~{\varepsilon_0 b L ^4\,{{V_\text{ES}}}^2}/{8\,{g}_0^{\frac{1}{2}}{E} {I}_{yy}\pi^4}$ is the voltage parameter, ${r}~=~\sqrt{{I}_{yy}/{A}}$ is the gyration radius of the cross section and ${N}_E~=~4\pi^2EI/L^2$ is the {Euler} buckling force of a straight beam. Eq.~\eqref{EQ:beamRO} shows that the deflection of the beam $w_m$ is parameterized by $\beta$ and by ${N}$, {where the latter} depends on the temperature and therefore on the flow velocity~\cite{kessler2016flow}.

The devices were fabricated using {silicon-on-insulator} substrates with $\approx 20\;\mu$m thick, highly doped, single crystal silicon device layer. Lithographically defined curved micromechanical beams were etched using deep reactive ion etching and released using hydrofluoric acid. Following release, the chip was glued to a custom built holder that was mounted onto a wafer probe station. The velocity of the pressure-controlled system was calibrated using a Pitot tube connected to a manometer with a resolution of {$\approx\,2.45$ Pa (0.01~in$\,H_2O$)}. During calibration, the end of the Pitot tube was set to measure the hydraulic head at the location of the chip. The velocity of the air stream was calculated using Bernoulli’s equation.

The beam deflection was measured using an optical microscope. We first applied a constant $V_\text{ET}\approx2$~V to induce Joule heating within the micromechanical beam. The actuation voltage $V_\text{ES}$ was then linearly increased from zero to $\approx 100$ V at a rate of $\approx 3$~V~s$^{-1}$. During this process, the motion of the beam was video recorded at the {frame rate of 10 s$^{-1}$}. The voltage-deflection curve was constructed using image processing techniques detailed in~\cite{kessler2016flow}. Next, the air flow was induced and the resulting beam response was measured at different flow velocities. In each case, $V_\text{ST}$ was extracted from the voltage-deflection curves.

\section{Results and discussion}\label{sec:results}

The voltage-displacement characteristics ($\beta=\beta(w_m)$) were obtained by solving Eq.~\eqref{EQ:beamRO} with $N$ calculated from Eq.~\eqref{EQ:N0} for $V_\text{ET} =2$ V. Device dimensions, used in the calculations, were measured using confocal microscopy. Due to the residual stress, the midpoint elevation of a released, "as fabricated" (at rest), beam differs from the nominal, ``as designed'', value $q_0$. This residual stress was estimated by solving Eq.~\eqref{EQ:beamRO} for $\sigma_r$  (with $\beta=0$ and $N$ given by Eq.~\ref{EQ:N0} with $ \overline{\theta} =0$) using the measured value $q_0\approx3.3~\mu$m. Our results show a stress value of $\sigma_r \approx 5.6 $ MPa \cite{Medina2014323}.

\begin{figure}[!ht]
\centering
\includegraphics[scale=0.4]{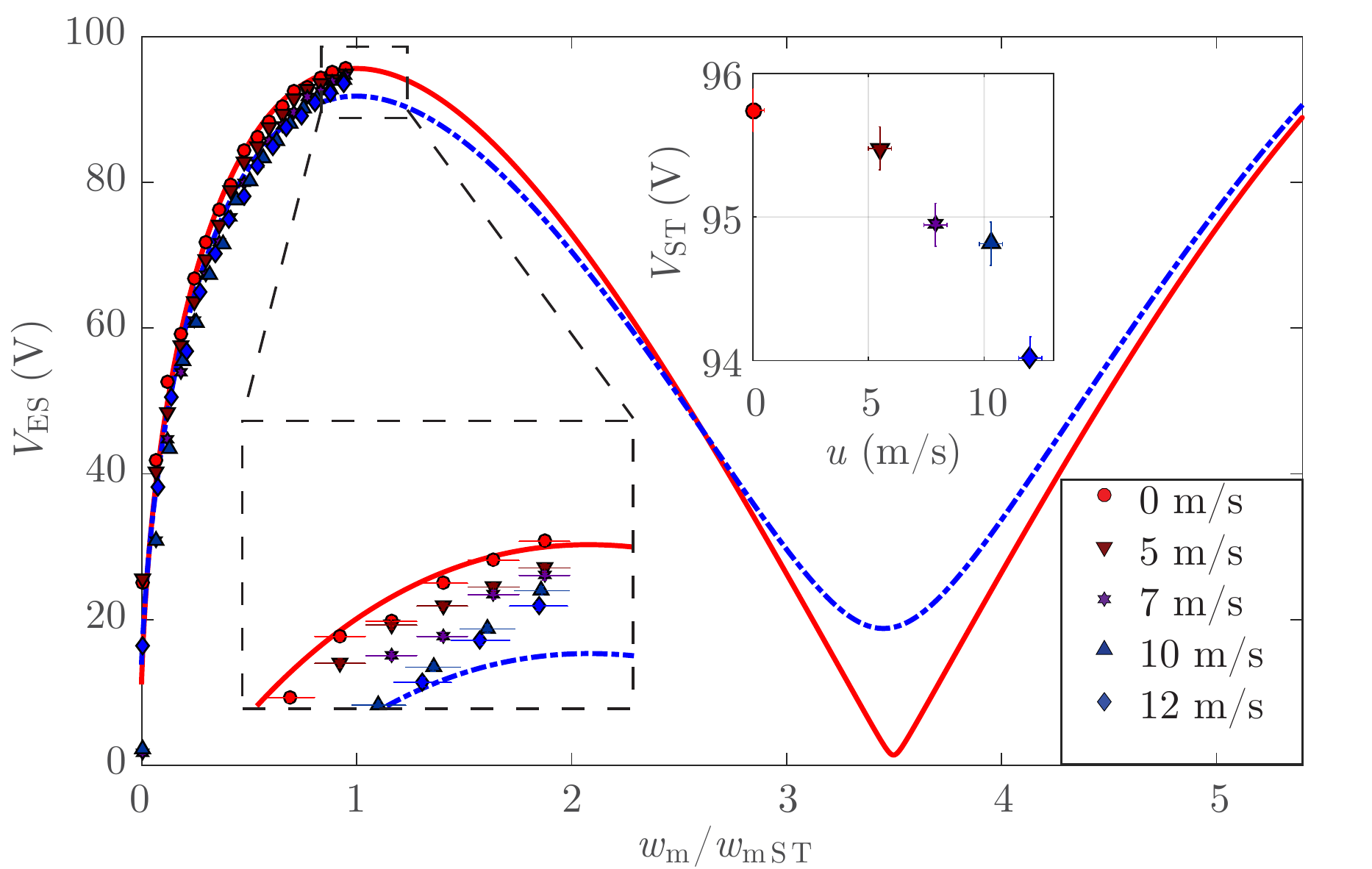}
\caption{Measured (markers) and calculated (lines) response of the beam to the electrostatic voltage ($V_{\text{ES}}$). The midpoint deflection $w_{m}$ of the beam is normalized by the snap-through value $w_{m ST}$ at each of the flow velocities $u$.  Solid and dashed lines represent the zero flow and the $u=12$~m$\,$s$^{-1}$ case, respectively. The lower left inset shows a zoomed-in view near the critical snap-through point. The upper right inset shows the experimental scale factor curve $V_\text{ST}=V_\text{ST}(u)$ of the flow sensor. {Error bars ($\pm 0.03, \,\pm 0.15 $~V$ $ and $ \,\pm 0.5 $~m$\,$s$^{-1}$ for the deflection, the voltage and the air-velocity, respectively) are obtained using uncertainty propagation analysis based on resolution errors of the measurement equipment, as described in the main text.}
}
\label{fig:w_Voltagebeta} \label{fig:w_Voltage}
\end{figure}

Results of calculations are shown in Fig.~\ref{fig:w_Voltagebeta} for zero flow and for an air flow velocity of $u=12$~m$\,$s$^{-1}$. Our data shows a decrease of $V_\text{ST}$ with increasing $u$. Experimental results, shown in Fig.~\ref{fig:w_Voltagebeta}, are consistent with the model predictions. The uncertainty in $V_\text{ES}$ and $V_\text{ST}$ of $0.3$~V is attributed to the time synchronization error, estimated to be one video frame or $\approx0.1$~s, between the video recording and the $V_\text{ES}$ signal. The accuracy of the flow velocity is limited by the resolution of the calibration tool, which is $1$~m$\,$s$^{-1}$.

As expected, at a certain voltage  ${V_\text{ES}} \approx V_\text{ST}$, corresponding to the limit (maximum) point of the equilibrium curve, the ST is observed when the beam jumps to a postbuckled configuration. Since the beam deflection was open-loop voltage controlled, only the stable branch of the equilibrium curve, up to ${V_\text{ES}}=V_\text{ST}$, can be obtained experimentally. The dependence of the measured $V_\text{ST}$ on $u$ (the scale factor curve $V_\text{ST}=V_\text{ST}(u)$) is shown in the upper inset of Fig.~\ref{fig:w_Voltagebeta}.
We define the device sensitivity as the slope of the scale factor curve. In the measured range of $u$, the sensitivity predicted by the model is $dV_{\text{ST}}/du  = 0.08$~V$\,$s$\,$m$^{-1}$ while the experimental value is $dV_{\text{ST}}/du \approx 0.13$~V\,s\,m$^{-1}$.

Our results show that the ST based sensor has lower power consumption than initially straight and then buckled beam~\cite{kessler2016flow}. Specifically, while the actuating voltage of the curved beam reported here is $V_\text{ET}\approx 2$~V
the Euler's buckling voltage of an identical straight beam, obtained by re-scaling the measured data from~\cite{kessler2016flow}, is $\approx 3$~V. For the measured $\approx11$~k$\Omega$ resistance of our curved beam, the power consumption was $\approx 0.36$~mW whereas for the straight beam with identical dimensions was $\approx 0.82$~mW.

In contrast to the Euler buckling, ST collapse is accompanied by an abrupt change of the beam configuration. As a result, the ST event is easily detectable and the critical ST voltage can be measured with high accuracy. Furthermore, utilization of integrated capacitive or optical sensing techniques will {bring} the device closer to practical implementation {as} a {gas} flow velocity and/or a wall shear stress sensor in real-life engineering applications.

\section*{Acknowledgment}
Devices were fabricated in part at the Center for Nanoscale Science and Technology (CNST) at the National Institute of Standards and Technology (NIST). The research is supported by the Israel Ministry of Science and Technology, grant 3-14411. The third author acknowledges support from the Henry and Dinah Krongold Chair of Microelectronics. The authors would like to thank Lior Medina for his help with the model.

\bibliographystyle{IEEEtran}
\bibliography{IEEEabrv,mybib}

\end{document}